# Nonlinear THz Control of the Lead Halide Perovskite Lattice

Maximilian Frenzel[1,*], Marie Cherasse[1,2,*], Joanna M. Urban[1], Feifan Wang[3,‡], Bo Xiang[3], Leona Nest[1], Lucas Huber[3,§], Luca Perfetti[2], Martin Wolf[1], Tobias Kampfrath[1,4], X.-Y. Zhu[3], Sebastian F. Maehrlein[1,†]

[1] Fritz Haber Institute of the Max Planck Society, Department of Physical Chemistry, Berlin, Germany
[2] LSI, CEA/DRF/IRAMIS, CNRS, Ecole Polytechnique, Institut Polytechnique de Paris, Palaiseau, France
[3] Columbia University, Department of Chemistry, New York City, USA
[4] Freie Universität Berlin, Berlin, Germany

* Authors contributed equally
‡ Present address: Department of Materials, ETH Zurich, 8093 Zürich, Switzerland
§ Present address: Sensirion AG, Staefa, Switzerland
† Corresponding author. Email: maehrlein@fhi.mpg.de

**Abstract**

**Lead halide perovskites (LHPs) have emerged as an excellent class of semiconductors for next-generation solar cells and optoelectronic devices. Tailoring physical properties by fine-tuning the lattice structures has been explored in these materials by chemical composition or morphology. Nevertheless, its dynamic counterpart, phonon-driven ultrafast material control, as contemporarily harnessed for oxide perovskites, has not been established yet. Here we employ intense THz electric fields to obtain direct lattice control via nonlinear excitation of coherent octahedral twist modes in hybrid $CH_3NH_3PbBr_3$ and all-inorganic $CsPbBr_3$ perovskites. These Raman-active phonons at 0.9 – 1.3 THz are found to govern the ultrafast THz-induced Kerr effect in the low-temperature orthorhombic phase and thus dominate the phonon-modulated polarizability with potential implications for dynamic charge carrier screening beyond the Fröhlich polaron. Our work opens the door to selective control of LHP's vibrational degrees of freedom governing phase transitions and dynamic disorder.**

Introduction

During the last decade, lead halide perovskites (LHPs) emerged as promising semiconductors for efficient solar cells, light-emitting diodes, and other optoelectronic devices (*1-3*). Key prerequisites for the high LHP device efficiencies are the long charge carrier diffusion lengths and lifetimes (*4, 5*), often explained by the unusual defect physics (*6, 7*) and/or dynamic charge carrier screening (*8, 9*). The latter relies on delicate electron-phonon coupling, established by the dominant role of the static structure and dynamics of the lead-halide framework (*10, 11*). However, the exact mechanisms of the carrier-lattice interaction in the highly polarizable and anharmonic LHP lattices remain debated (*12, 13*). The sensitivity of the physical properties to structural distortions is a common feature in the extensive family of perovskites. In particular, for oxide perovskites, the control of specific lattice modes leads to ultrafast material control and nonlinear phononics (*14, 15*). Successful examples include, among others, light-induced superconductivity (*16*), magnetization switching (*17*), access to hidden quasi-equilibrium spin states (*18*), ferroelectricity (*19, 20*) and insulator-metal transitions (*21*) in perovskite or similar garnet structures.

The crystal structure of LHPs features a large A-site cation surrounded by $PbX_6$ octahedra consisting of lead (Pb) and halide (X) ions in the $ABX_3$ crystal structure (see **Fig. 1A**). The electronic band structure is mainly determined by the identities of metal and halide but is also highly sensitive to the Pb-X-Pb bond angle, which can be controlled through the steric hindrance of the A-cation (*22*). Changing the Pb-X-Pb bond angle is equivalent to tilting of the PbX$_6$ octahedra, which serves as an order parameter for the cubic → tetragonal → orthorhombic phase transitions (*23, 24*). Octahedral tilting is also an important factor governing structural stability (*25*), dynamic disorder (*26, 27*), and potential ferroelectricity (*26, 27*) in LHPs. A recent study using resonant excitation of the ~1 THz octahedral twist mode (Pb-I-Pb bending) revealed modulation of the bandgap of CH$_3$NH$_3$PbI$_3$ at room temperature (*28*). A similar observation of dynamic bandgap modulation due twist modes was made at 80K for off-resonant impulsive Raman excitation (*29*). These twist modes are also believed to contribute to the formation of a polaronic state (*30*). All of these findings indicate an intriguing role of carrier coupling to Raman-active non-polar phonons in addition to the polar LO phonons in the conventional Fröhlich polaron picture (*11, 31*). In addition, the application of the Fröhlich polaron picture to LHPs has been questioned (*9, 26*), because of the limited applicability of the harmonic approximation in these soft lattices (*13*).

Accordingly, the dynamic screening picture in LHPs is incomplete and its microscopic mechanism continues to be debated (*32, 33*). Furthermore, identifying and characterizing polaronic behavior is experimentally difficult (*31, 33-37*). Optical Kerr effect (OKE) in LHPs (*38, 39*) did not succeed in unveiling a lattice response and can be explained by an instantaneous electronic polarization (due to hyperpolarizability) instead (*40*). Moreover, previous strong field THz excitation could not directly detect the driven vibrational modes (*28, 31*) and coherent control of the phonons remained elusive. Here, we turn to the THz-induced Kerr effect (TKE) (*41, 42*) to investigate lattice-modulated polarization dynamics in the electronic ground state. We employ intense THz electric fields (**Fig. 1B**) that broadly cover most of the inorganic cage modes (**Fig. 1C**) and may nonlinearly probe the THz polarizability. The rapidly changing single-cycle THz field macroscopically mimics the sub-picosecond variation of local electric fields following electron-hole separation (*43, 44*) and elucidates the isolated lattice response.

**Experiment**

Generally, the polarizability describes the tendency of matter to form an electric dipole moment when subject to an electric field, such as the local field from a mobile charge carrier in a semiconductor. In the presence of an electric field $\boldsymbol{E}$, the microscopic dipole moment is given by $\boldsymbol{p}(\boldsymbol{E}) = \boldsymbol{\mu}_0 + \alpha \boldsymbol{E}$, where $\mu_0$ is the static dipole moment and $\alpha$ is the polarizability tensor. In LHPs $\alpha$ originates from three contributions: instantaneous electronic response ($\alpha_\text{e}$), lattice distortion ($\alpha_\text{lat}$), and molecular A cation reorientation ($\alpha_\text{mol}$). For small perturbations of the respective collective coordinate $Q$ (charge distribution, molecular orientation, or lattice mode) a Taylor expansion yields

$$\boldsymbol{p}(\boldsymbol{E}, Q) = \boldsymbol{\mu}_0 + \frac{\partial \boldsymbol{\mu}_0}{\partial Q} Q + \frac{\partial \alpha}{\partial Q} Q \boldsymbol{E}, \quad (1)$$

where the two partial derivatives correspond to the mode effective charge $Z^*$ and the Raman $R_{ij}$ tensor, respectively. Macroscopically, the two terms lead to lattice polarization $Z^* Q_\text{IR}$ and phonon-modulated susceptibility $\chi^{(1)}_\text{eq} + (\partial \chi^{(1)}_\text{eq}/\partial Q_\text{R}) Q_\text{R}$ for polar, $Q_\text{IR}$, and non-polar, $Q_\text{R}$, modes, respectively. The latter relates $\partial \alpha / \partial Q$ to a transient dielectric function and change in refractive index of the material. This relation thus enables studying microscopic polarizability through the observation of macroscopic transient birefringence induced by a pump pulse and experienced by a weak probe pulse (*41, 45*). Collective polarization dynamics are induced by the driving force $F = -\partial W_\text{int}/\partial Q$, where $W_\text{int} = -\boldsymbol{P}(\boldsymbol{E}, Q) \cdot \boldsymbol{E}$ is the potential energy of the macroscopic polarization $\boldsymbol{P} = \sum_i \boldsymbol{p}_i$ interacting with an electric field $\boldsymbol{E}$ (from a local charge carrier or through light-matter coupling in the electric dipole approximation). Thus, two $E_{\text{THz}}$ interactions lead to THz polarizability-induced transient birefringence in TKE (*42*), which is linearly probed by a weak probe pulse $E_{\text{pr}}$ in an effective 3$^{\text{rd}}$ order nonlinear process proportional to $\chi^{(3)} E_{\text{THz}} E_{\text{THz}} E_{\text{pr}}$ (see **Methods**) (*41, 46*).

To induce polarization dynamics, we use intense THz single-cycle pulses with a 1.0 THz center frequency (> 1.5 THz spectral width, see **Fig. 1C**), delivering sub-picosecond peak electric fields exceeding 1.5 MV/cm generated by optical rectification in LiNbO$_3$ (*47*). We probe the resulting transient birefringence, i.e. anisotropic four-wave mixing signals, by stroboscopic sampling with a synchronized 20 fs pulse (800 nm center wavelength) in a balanced detection scheme, see **Fig. 1A**. We therefore effectively measure a 3$^{\text{rd}}$ order nonlinear signal field heterodyned with the transmitted probe field. The probe pulses are linearly polarized at 45° with respect to the vertically polarized THz pulses. As representative LHPs, we investigate hybrid organic-inorganic CH$_3$NH$_3$PbBr$_3$ (MAPbBr$_3$) and fully inorganic CsPbBr$_3$. The freestanding single crystal samples (200 – 500 µm thickness) were solution grown by an antisolvent diffusion method (*48, 49*) (see **Methods**). Complementary polycrystalline thin films (~ 400 nm thickness) were spin-coated on 500 µm BK7 substrates, being particularly technologically relevant as most state-of-the-art LHP solar cells are fabricated in a similar way (*50*).

**Results**

**Fig. 2A** shows the THz induced transient birefringence in MAPbBr$_3$ single crystals at room temperature. The signal (blue line) initially follows $E_{\text{THz}}^2$ (grey area, measured via electrooptic sampling), but then transitions into a nearly mono-exponential decay for time delays $t >$ 500 fs. The transient birefringence peak at $t = 0$ clearly scales quadratically with the THz-field amplitude as found by the pump fluence dependence in **Fig. 2B**. With the exponential decay dynamics remaining also constant for different fluences (**Fig. S2**), we can infer the Kerr-type origin of the full signal and thus conclude a strong THz polarizability. Furthermore, the peak amplitude's (**Fig. 2C**) and the exponential tail's (**Fig. S3**) dependence on the azimuthal angle between probe polarization direction and crystal axes perfectly obeys the expected 4-fold rotational symmetry of the $\chi^{(3)}$ tensor and TKE dependence of $\chi^{(3)}_{ijkl} E^{\text{THz}}_j E^{\text{THz}}_k E^{\text{pr}}_l$. We quantify the THz polarizability of MAPbBr$_3$ by a nonlinear THz refractive index $n_2$ of about $2 \times 10^{-14}$ cm$^2$/W (see details in SI), being on the same order as in the optical region (*51*) and roughly 80 times larger than $n_2$ of Diamond (*52*), which is known as a suitable material for THz nonlinear optics (*53*).

The small oscillatory deviations from the exponential tail in MAPbBr$_3$ (**Fig. 2A**), become more pronounced and qualitatively different in CsPbBr$_3$ in the form of a bumpy, non-trivial shape (**Fig. 2D**). This stark difference between MAPbBr$_3$ and CsPbBr$_3$ is reminiscent of 2D-OKE results (*40*), where the oscillatory signal of CsPbBr$_3$ was found to be mainly due to anisotropic light propagation, since CsPbBr$_3$ is orthorhombic and thus birefringent at room temperature. The fluence (**Fig. 2E**) and azimuthal dependences (**Fig. 2F**) are consistent with the pure 3$^{\text{rd}}$ order nonlinearity of the signal. However, fits to the azimuthal angle dependences in **Figs. 2C,F** (black lines) yield different ratios of the off-diagonal $\chi^{(3)}_{ijkl}$ to diagonal $\chi^{(3)}_{iiii}$ tensor elements for the two materials: 1.6 for MAPbBr$_3$ and 1.0 for CsPbBr$_3$. A similar polarization dependence of static Raman spectra was recently attributed to additional isotropic disorder from the rotational freedom of the polar MA$^+$ cation in MAPbBr$_3$ (*54*).

**Figs. 3A,B** show a comparison of the temperature dependent TKE in MAPbBr$_3$ single crystals and polycrystalline thin films. At room temperature (both top traces), it stands out that the thin film TKE signal lacks the exponential decay seen in the single crystals, providing a first evidence that the tail stems from dispersion effects and is not due to intrinsic molecular relaxation dynamics as previously interpreted (*55*). A strong and sophisticated THz dispersion, as seen in **Fig 1C**, is a general, but often overlooked, phenomenon in broadband high-field THz pump-probe spectroscopy. In analogy to the OKE (*40*), the features of the room temperature TKE in both MAPbBr$_3$ and CsPbBr$_3$ might therefore be dominated by dispersive and anisotropic light propagation. Hence, we assign the main contribution of the TKE response at room temperature to the instantaneous electronic polarizability (hyperpolarizability), which may overwhelm possible lattice contributions. This interpretation will be further supported by the modeling below.

From here on, we mainly focus on the TKE of MAPbBr$_3$, especially at low temperatures at which increased phonon lifetimes should facilitate the observation of a coherent lattice response (*54, 56*). For the single crystal (**Fig. 3A**), the TKE dynamics at 180 K are different than at room temperature, which might reflect the change of structural phase from cubic to tetragonal. At 180K, an oscillatory signal at short times (< 2 ps) appears, suggesting the presence of a coherent phonon which was overdamped in the cubic phase at room temperature (*54*). The coherent oscillations become much stronger for the single crystal at 80 K, where MAPbBr$_3$ is in the orthorhombic phase. Less pronounced, but clear oscillations are also visible in the thin film sample at 80 K (**Fig. 3A**, lowest trace). We extract the oscillatory parts, **Fig. 3C**, of both single crystal and thin film samples at 80 K by subtracting incoherent backgrounds, using a convolution of the squared THz field with a bi-exponential function. The respective Fourier transforms in **Fig. 3D** reveal the same oscillations frequency of 1.15 ± 0.05 THz for both samples. This clearly rules out anisotropic propagation effects as the origin of these oscillations (*40*), as the 400 nm film is too thin for significant walk-off between pump and probe (shown in simulations later) and the different thicknesses of the two samples rule out a Fabry-Pérot resonance effect. Thus, we can clearly assign the signal to a lattice-modulation of the THz polarizability dominated by a single 1.15 THz phonon in MAPbBr$_3$. We now turn to THz-THz-VIS four-wave-mixing simulations to understand the origins of TKE from MAPbBr$_3$.

**Modelling**

For dispersive and birefringent materials, the Kerr signal cannot be decomposed into an effective birefringence change observed by an independent probe beam (*46*). Instead, the Kerr-effect induced nonlinear polarization $P^{(3)}$ needs to be captured in a full four-wave-mixing (FWM) picture. To separate the three polarizability contributions (instantaneous electronic, molecular and lattice) and to take anisotropic light propagation across dispersive phonon resonances into account, we simulate the 3$^{\text{rd}}$ order nonlinear polarization by

$$P_i^{(3)}(t,z) = \epsilon_0 \int_{-\infty}^{t} dt' \int_{-\infty}^{t'} dt'' \int_{-\infty}^{t''} dt''' \, \tilde{R}_{ijkl} R(t,t',t'',t''') E_j^{\text{THz}}(t',z) E_k^{\text{THz}}(t'',z) E_l^{\text{pr}}(t''',z), \quad (2)$$

where $R$ is the time-domain $\chi^{(3)}$ response function (*46*) and $E^{\text{THz}}$ and $E^{\text{pr}}$ are the pump and probe electric fields, respectively. The time-independent $\tilde{R}_{ijkl}$ tensor constitutes the respective $\chi^{(3)}$ symmetry for the different crystalline phases, in agreement with the ratios of the tensor elements obtained from the azimuthal fits in **Fig. 2C**. For the instantaneous electronic polarizability (hyperpolarizability), we assume temporal Dirac delta functions $R_e(t,t',t'',t''') = R_{e,0}\delta(t-t')\delta(t'-t'')\delta(t''-t''')$. For a lattice response, we model the driven phonon response by a Lorentz oscillator

$$R_{\mathrm{ph}}(t,t',t'',t''') = R_{\mathrm{ph},0}\delta(t'-t'')\delta(t''-t''')e^{-\Gamma(t-t')}\sin\left[\sqrt{(\omega_{\mathrm{ph}}^2-\Gamma^2)}(t-t')\right], \quad (3)$$

where $\omega_{\mathrm{ph}}/2\pi$ is the frequency and $1/2\Gamma$ the lifetime of the phonon (*46*). The driving force for Raman-active phonons is hereby $E_j^{\mathrm{THz}}E_k^{\mathrm{THz}}$, which contains difference- and sum-frequency terms (*57, 58*). The latter is a unique distinction to the OKE. For $E^{\mathrm{THz}}$ we can directly use the experimental THz electric field, as measured in amplitude and phase resolved electro-optic sampling. After we determine the complex refractive indices (**Fig. 1C**) and extrapolate the static birefringence (see Methods and SI), we calculate and propagate all involved fields from **Eq. (2)** incl. signal fields $E_i^s(t,z)$ emitted from $P_i^{(3)}(t,z)$, followed by our full detection scheme, including balanced detection, to obtain the pump-probe signal (see details in **Methods**).

**Fig. 4A** shows the simulated TKE signal (grey) compared to the experimental data (blue) at room temperature for a 500 μm thick MAPbBr3 single crystal. It unveils the formation of a long exponential tail produced by walk-off, dispersion, and absorption effects, even for only an instantaneous electronic response $R_e$. This confirms that the electronic polarizability dominates the TKE signal at room temperature. It contrasts a previous interpretation of a TKE measurement in thick single crystal MAPbBr3, which neglected propagation effects entirely (*55*). At 80 K, MAPbBr3 is orthorhombic. We therefore need to include additional static birefringence. Instantaneous hyperpolarizability alongside static birefringence and dispersion can cause the appearance of oscillatory features (*40*). Nevertheless, our modelling finds these features to be too short-lived (see **Fig. S14**) to explain our experimental observation at 80 K. Thus, we need to account for both hyperpolarizability $R_e$ and lattice-modulated polarizability $R_{\mathrm{ph}}$ responses (fit parameters: $\omega_{\mathrm{ph}}/2\pi$ = 1.14 THz, $\Gamma = (2\cdot 1.7\mathrm{ps})^{-1}$, $R_{e,0}/R_{\mathrm{ph},0}$ = 2.4) to describe the low-temperature TKE signals in the time- and frequency domain (**Figs. 4B,C**). In contrast to OKE at 80K (*40*), the oscillations in TKE are therefore due to coherent phonon modes and we hence finally observe an ultrafast lattice response to a sub-picosecond electric field transient.

The simulation assuming only instantaneous hyperpolarizability for a 400 nm thin film agrees well with the experimental TKE at room temperature (see **Fig. 4D**). As expected, the simulation lacks the clear tail seen in the thick single crystals, thereby additionally confirming that the tail is due to light propagation effects. Also here, at 80 K, we need to include both instantaneous electronic and phonon contributions ($\omega_{\mathrm{ph}}/2\pi$ = 1.14 THz, $\Gamma = (2\cdot 1.7\mathrm{ps})^{-1}$, $R_{e,0}/R_{\mathrm{ph},0}$ = 24) to describe the experimental signals for the thin films in **Figs. 4E,F**. Here, a purely instantaneous electronic contribution alongside static birefringence does not lead to oscillatory features (see **Figs. S14A,C**). This provides direct proof that the observed oscillations in **Figs. 3C,D** originate from a coherent phonon. Therefore, through comparison of single crystals with thin films and by rigorous FWM simulation, we prove to witness a coherent lattice-driven dynamic polarization response.

**Interpretation**

Besides potential rotational disorder, our rigorous modeling shows that we do not observe a TKE contribution that we can unambiguously relate to an ultrafast cation reorientation in the form of a liquid-like exponential decay (*41, 42*). We rather find MAPbBr3's TKE tail at room temperature to be most likely overwhelmed by the instantaneous hyperpolarizability $R_e$ in conjunction with dispersive light propagation. This might be also explained by the THz pump spectrum being far off the cation rotational resonances around the 100 GHz frequency range (*59*). The cation species nevertheless influences the static and dynamic properties of the inorganic lattice, highlighting the importance of the interplay between the organic and inorganic sub-lattices for the LHPs equilibrium structure (*56*). This fact shows up e.g. as a single dominating PbBr$_6$ cage mode in MAPbBr$_3$ but two dominating modes in CsPbBr$_3$ (see **Fig. S1**); in agreement with static Raman spectra (*54*). The various templating mechanisms by which the cation influences these properties (*60*) are through its steric size (*22*), lone-pair effects (*27, 61*), or hydrogen bonding (*62*).

For MAPbBr$_3$, we find a single phonon mode dominating the Raman-active lattice dynamics in response to a sub-ps electric field spike. The observed phonon at 1.15 THz is consistent with static Raman spectra in the visible range, where this mode also exhibits the highest scattering amplitude (*54, 63*). Thus, we can assign it to a dynamic change in the Pb-Br-Pb bond angle corresponding to a twisting of the PbBr$_6$ octahedra (twist mode) (*64*). Based on theory work for MAPbI$_3$ (*65*), we assign this to A$_g$ symmetry, which matches the experimental observations that the mode is still present when we rotate the single crystal by 45° (see **Fig. S7**) and that we also observe the same mode in polycrystalline thin films (**Figs. 3C,D**). We suggest that at room temperature this mode also strongly modulates the THz dielectric response, even though its oscillations are potentially overdamped as inferred from the broad Raman spectra (*54, 56*). To distinguish whether this twist mode only dominates the ultrafast lattice response in MAPbBr$_3$, or is of wider relevance for other LHPs, we analyze the TKE response of CsPbBr$_3$, where we observe two modes at 0.9 and 1.3 THz at 80K (see **Fig. S1**), corresponding to two octahedra twist modes as observed in static Raman spectra (*54*). We thus conclude that the transient THz polarizability $(\partial \chi_{eq}^{(1)}/\partial Q)\, Q$ is generally dominated by the octahedra twist modes in LHPs.

We now consider the excitation mechanism of the coherent phonon. **Fig. 5A** shows that the 1.15 THz oscillations at 80K scale with the square of the THz electric field amplitude, suggesting nonlinear excitation with a Raman-type driving force. This is consistent with the Kerr effect being also a Raman-type probing mechanism. Generally, there are four types of Raman-active THz excitation mechanisms: Difference- or sum-frequency excitation via Ionic Raman Scattering (IRS) or Stimulated Raman Scattering, corresponding to nonlinear ionic (=phononic) or nonlinear electronic (= photonic) pathways, respectively (*58*). Indeed, the A$_g$ symmetry of the observed modes permits IRS, where a resonantly driven IR-active phonon couples anharmonically to a Raman-active mode (*14, 58*). However, this phononic pathway requires phonon anharmonicity, whereas the photonic pathway requires electronic THz polarizability. The sum-frequency (SF) and difference-frequency (DF) photonic force spectra in **Fig. 5B** indicate a comparable probability for both photonic mechanisms to drive the 1.15 THz mode (dashed line). For the phononic pathways in **Fig. 5C**, the DF excitation requires a primarily driven IR-active phonon with a bandwidth of ≳ 1 THz, which exists in our excitation range even at 80 K (*66*). On the other hand, there are also IR-active modes, which provide roughly half the frequency of the Raman-active mode $\Omega_{IR} = \Omega_R/2$ enabling phononic SF-IRS (*58*). Accordingly, none of the four nonlinear excitation pathways can be neglected, but the observed strong electronic THz polarizability in conjunction with a longer penetration depth for lower THz frequencies favors a SF nonlinear photonic mechanism. We leave the determination of the exact excitation pathway to further studies, e.g. by two-dimensional TKE (*67*) or more narrowband THz excitation (*68*).

**Discussion**

Independent of the precise excitation pathway and in contrast to optical Raman or transient absorption studies, we unambiguously observe strong electron-phonon coupling of the octahedral twist modes via a pure THz polarizability (electronic or ionic). This explains the mode's dominating influence on the electronic bandgap in MAPbI$_3$ previously observed by Kim et al. (*28*). The twist mode's half-cycle period of ~0.5 ps is short enough to contribute to electron-phonon coupling within the estimated polaron formation time (*69*), even in the overdamped case at room temperature. We can understand carrier screening by non-polar modes as follows. As shown in **Eq. (1)**, the THz polarizability contains two lattice contributions: From polar lattice modes $P_{IR}(\omega) \propto Z^* Q_{IR}(\omega) \propto Z^* E_{THz}(\omega)$ and from the non-resonant electron cloud moving at THz speeds (sub-ps time scales):

$$P_e(\omega) = \epsilon_0 [\, \chi_e^{(1)}(\omega) + \frac{\partial \chi_e^{(1)}}{\partial Q_R}(\omega, \Omega) \, Q_R(\Omega) \,] E_{THz}(\omega), \qquad (4)$$

where the latter is modulated in the presence of a Raman-active phonon $Q_R$. Thus, excited Raman-active modes lead to a transient dielectric response $\epsilon(\omega) = \epsilon_{eq}(\omega) + \Delta\epsilon(\omega, \Omega)$ at THz frequencies $\omega$ with $\Delta\epsilon = \frac{\partial \chi_e^{(1)}}{\partial Q_R} Q_R$, which constitutes an additional contribution of higher order screening due to a fluctuating lattice. In the macroscopic incoherent case, $\Delta\epsilon$ averages out. On time and length scales relevant to electron-hole separation and localization (< 1 nm and < 1ps) (*43, 44*), collective octahedral tilting produces (*70*) an additional THz polarizability, which might add to the conventional Fröhlich picture of carrier screening. We speculate that a local non-zero twist angle $Q_R$ could be either already present due to dynamic disorder (see discussion below), or might be nonlinearly excited by the transient local charge field $E_{loc}^2$, easily exceeding 1 MV/cm (*9*) (analog to the excitation pathways above). The latter scenario agrees with MAPbBr$_3$'s unusually large optical $\chi^{(3)}$, previously attributed to local confinement effects (*51*). The observed 1.15 THz mode is therefore a good candidate for contributing to strong electron-phonon coupling beyond the polar Fröhlich picture.

The driven twist mode is similar to soft modes in oxide perovskites, where the tilting angle of adjacent oxygen octahedra is an order parameter for phase transitions (*71*). Recently, TKE was similarly employed to drive and detect ultrafast field-induced ferroelectricity in the quantum paramagnet SrTiO$_3$ (*19*). In Eu and Sr doped La$_2$CuO$_4$, driving the tilt of oxygen octahedra was found to induce signatures of superconductivity persisting over a few ps above the critical temperature (*16*). Consistent with these observations in oxide perovskites, the tilting angle of the PbX$_6$ octahedra (twist mode) was found to act as an order parameter for phase transitions in LHPs (*23, 24*) and in the double-perovskite Cs$_2$AgBiBr$_6$ (*72*). Especially for MAPbBr$_3$, the Raman scattering intensity of the 1.1 THz peak was recently shown as measure of the orthorhombic-tetragonal phase transition (*63*) and its spectral evolution in Raman (*56*) and neutron scattering (*73*) is indicative of a soft mode phase transition. Yet, the LHP lattice properties were previously mainly tuned in a static and chemical manner, e.g. by acting on the octahedral tilting angle through the steric size of the A-site cation (*22*). The coherent lattice control demonstrated here allows dynamic tuning of the structure and thus ultrafast phonon-driven steering of LHP's optoelectronic properties, e.g. for integrated photonic devices operating at GHz to THz clock-rates (*74*).

In addition, imposing a coherence on the octahedral tilting should directly influence the dynamic disorder (*75*), which is considered one of the key components determining the optoelectronic properties of LHPs (*12, 54, 76*). Dynamic disorder means that the effective crystallographic structure (e.g. cubic at 300 K) only exist in spatial and temporal average. Specifically, in LHPs with a Goldschmidt tolerance factor below 1, such as MAPbBr$_3$ and CsPbBr$_3$, the disorder mainly arises from the lattice instability associated with octahedral tilting (*61, 75, 77*), evidenced by X-ray total scattering in CsPbBr$_3$ (*78*), inelastic X-ray scattering in MAPbI$_3$ (*70*) and Raman spectroscopy in MAPbBr$_3$, CsPbBr$_3$ and MAPbI$_3$ (*54, 77*). The resulting fluctuating lattice potential and polar nanodomains have been suggested as underlying mechanisms for dynamic charge carrier screening in the form of preferred current pathways (*79, 80*) and ferroelectric polarons (*26, 81*), respectively. All these phenomena might be potentially controlled or temporally lifted by the THz control of octahedral motion.

Overall, we find that the octahedral tilting motion, which serves as an order parameter for phase transitions (*23, 24*) and contributes significantly to dynamic disorder (*54, 77*), shows a strong nonlinear coupling to a rapidly varying electric field on sub ps-timescales that are relevant to local electron-hole separation polaron formation. Our results thus indicate that the TO octahedral twist mode contributes to strong electron-phonon coupling and dynamic carrier screening in LHPs, which may be inherently linked to a local and transient phase instability as suggested by the ferroelectric polaron picture (*26, 81*).

**Conclusion**

By investigating 3$^{rd}$ order nonlinear polarization dynamics in hybrid and all-inorganic LHPs, we reveal that the room temperature TKE response stems predominantly from a strong THz hyperpolarizability, leading to a nonlinear THz refractive index on the order of 10$^{-14}$ cm$^2$/W. In analogy to previous OKE studies (*40*), we explain and model the appearance of retarded TKE dynamics by dispersion, absorption, walk-off, and anisotropy effects (*46*). These effects are of crucial relevance to contemporary THz pump-probe experiments, such as TKE or THz-MOKE studies (*82, 83*). For sufficiently long phonon lifetimes at lower temperatures, we can nonlinearly drive and observe a coherent lattice response of the ~1 THz octahedral twist mode(s). These phonons couple most strongly to the THz polarizability, which means they must be highly susceptible to transient local fields on the 100s fs time scale, relevant to electron-phonon coupling and carrier localization. We find this ultrafast non-polar lattice response to be mediated by anharmonic phonon-phonon coupling and/or by the strong nonlinear electronic THz polarizability. The same octahedral twist mode serving as a sensitive order parameter for structural phase transitions (*63, 73*) is likely the origin of significant intrinsic dynamic disorder in LHPs (*54, 75*). Thus, our findings suggest that the microscopic mechanism of the unique defect tolerance (*39, 84*) and long carrier diffusion lengths (*4, 5*) of LHPs might also rely on small phase instabilities accompanying the polaronic effects.

Our work demonstrates the possibility of coherent control over the twist modes via nonlinear THz excitation. Since the octahedral twist modes are the dynamic counterparts to steric engineering of the metal-halide-metal bond angle, our work paves the way to study charge carriers in defined modulated lattice potentials, to control dynamic lattice disorder, or to macroscopically switch polar nanodomains leading to the emergence of transient ferroelectricity.

## Materials and Methods

### Sample Growth
The single crystal samples were synthesized based on our previous published method (*40*). For MAPbBr$_3$, the precursor solution (0.45 M) was prepared by dissolving equal molar ratio of MABr (Dyesol, 98%) and PbBr$_2$ (Aldrich, ≥98%) in dimethylformamide (DMF, Aldrich, anhydrous 99.8%). After filtration, the crystal was allowed to grow using a mixture of dichloromethane (Aldrich, ≥99.5%) and nitromethane (Aldrich, ≥96%) as the antisolvent (*48*). Similar method was used for CsPbBr$_3$ crystal growth (*49*). The precursor solution (0.38 M) was formed by dissolving equal molar ratio of CsBr (Aldrich, 99.999%) and PbBr$_2$ in dimethyl sulfoxide (EMD Millipore Co., anhydrous ≥99.8%). The solution was titrated by methanol till yellow precipitates show up and did not redissolve after stirring at 50 °C for a few hours. The yellow supernatant was filtered and used for the antisolvent growth. Methanol was used for the slow vapor diffusion. All solid reactants were dehydrated in a vacuum oven at 150 °C overnight and all solvents were used without further purification.

Thin films. Before spin-coating, the substrate was rinsed by acetone, methanol and isopropanol and treated under oxygen plasma for 10 min. The freshly prepared substrate was transferred to the spin coater within a short time. For MAPbBr$_3$, the precursor DMSO (Aldrich, ≥99.9%) solution (2M) containing the equimolar ratio of MABr and PbBr$_2$ was used for the one-step coating method. The film was formed by spin-coating at 2000 rpm for 45 s and annealed at 110 °C for 10 min. For CsPbBr$_3$, a two-step method was implemented. First, the PbBr$_2$ layer was obtained by spin-coating the 1 M PbBr$_2$/DMF precursor solution at 2000 rpm for 45 s and dried at 80 °C for 30 min. Subsequently, the PbBr$_2$ film was immersed in a 70 mM CsBr/methanol solution for 20 min. Following the rinsing by isopropanol, the film was annealed at 250 °C for 5 min to form the uniform perovskite phase.

### THz-induced Kerr effect
THz pulses with 1.0 THz center frequency and field strength exceeding 1.5 MV/cm (**Fig. 1B,C**), were generated by optical rectification in LiNbO$_3$ with the tilted pulse front technique (*47*). To that end, LiNbO$_3$ was driven by laser pulses from an amplified Ti:sapphire laser system (central wavelength 800 nm, pulse duration 35 fs FWHM, pulse energy 5 mJ, repetition rate 1 kHz). The probe pulses came from a synchronized Ti:sapphire oscillator (center wavelength 800nm, repetition rate 80 MHz) and were collinearly aligned and temporarily delayed with respect to the THz pulse. The probe polarization was set at 45 degrees with respect to the vertically-polarized THz pulses. The THz pulses induced a change in birefringence (TKE) in the sample (*41*). This birefringence causes the probe field to acquire a phase difference between polarization components parallel and perpendicular to THz pulse polarization. The phase difference is detected via a half- and quarter-waveplate (HWP and QWP) followed by a Wollaston prism to spatially separate perpendicularly polarized probe beam components. The intensity of the two beams is detected by two photodiodes in a balanced detection configuration.

### Four-wave-mixing simulation
The 3$^{rd}$ order nonlinear polarization $P^{(3)}(t,z)$ is simulated using the general four-wave mixing equation (**Eq. (2)**) and according to Ref. (*46*). To compute $P^{(3)}(t,z)$, all three contributing light fields, $E_j^{\text{THz}}$, $E_k^{\text{THz}}$ and $E_l^{\text{pr}}$, are propagated through the crystal on a time-space grid. The three fields inside the sample are calculated at any location $z$ using

$$E_i(t,z) = \int_{-\infty}^{\infty} t_i(\omega) A_i(\omega) e^{-i(\omega t - k_i(\omega)z)}(1 - R_i(\omega,z))d\omega, \quad (5)$$

with

$$R_i(\omega, z) = r_i\left(1 + e^{2izk_i(\omega)}\right)\frac{e^{2i(d-z)k_i(\omega)}}{1 - r_i^2(\omega)e^{2idk_i(\omega)}}, \quad (6)$$

where $A_i(\omega)$ is the spectral amplitude of the field and $t_i$ and $r_i$ denote the Fresnel transmission and reflection coefficients respectively. As the input pump field $E^{\text{THz}}$ we use the full experimental THz electric field generated via optical rectification in LiNbO$_3$ as measured using electro-optic sampling in Quartz (*85*). For the probe field $E^{\text{pr}}$ we assume a Fourier limited Gaussian spectrum with center wavelength 800 nm and pulse duration 20 fs, experimentally measured by a spectrometer and a commercial SPIDER. For both non-birefringent and birefringent simulations, we use the THz refractive index for MAPbBr$_3$ as calculated from its dielectric function based on the experimental work by Sendner et al. (*10*) (**Fig. S11**). In the optical region, the precise anisotropic refractive index of CsPbBr$_3$ is used as measured using the 2D-OKE (*46*). For the birefringent lead halide perovskite simulation, the static birefringence of CsPbBr$_3$ is used and interpolated to THz region (**Fig. S12**). For the isotropic cubic perovskite, the static birefringence is set to zero. In the shown simulation results, the time-grid had a finite element size of $\Delta t' = 16.6$ fs and the spatial grid had a finite element size of $\Delta z = 10$ μm for the single crystal and $\Delta z = 0.1$ μm for the thin film simulations respectively. These values were chosen for the sake of computational efficiency and did not qualitatively affect the simulation results. The pump-probe delay finite element size was chosen to be $\Delta t = 16.6$ fs.

We assume that the nonlinear polarization $P^{(3)}$ emits an electric field $E^{(4)}$ at every slice $z$ according to the inhomogeneous wave equation

$$[\nabla^2 + k_i^2(\omega)]E_i^{(4)}(\omega, t, z) = -\frac{\omega^2}{\epsilon_0 c^2}P_i^{(3)}(\omega, t, z), \quad (7)$$

which then co-propagates with the probe field $E^{\text{pr}}$. The transmitted probe field $E^{\text{pr}}$ and emitted field $E^{(4)}$ are projected on two orthogonal polarization components by propagating through a half-wave plate, quarter-wave plate and Wollaston prism. The combined effect of these optical devices are captured by the Jones matrices $J_1$ and $J_2$ for the two separated polarization component channels. A balanced detection scheme allows observation of $E^{(4)}$ by interfering with $E^{\text{pr}}$. Under balanced conditions, the detected non-equilibrium signal is

$$S \propto \int Re\left[(J_1 E^{pr}) \cdot (J_1 E^{(4)*}) - (J_2 E^{pr}) \cdot (J_2 E^{(4)*})\right]d\omega. \quad (8)$$

Our simulation therefore mimics the balancing conditions of the experiment. A detailed description of this calculation is given in (*46*).

To model the response of the system, we assume the response function $R_e(t, t', t'', t''')$ for an instantaneous electronic response and $R_{\text{ph}}(t, t', t'', t''')$ for a phonon response. The expressions for $R_e$ and $R_{\text{lat}}$ are given in the main paper. In the frequency domain, $R(\omega) = \chi^{(3)}(\omega)$. For normal incidence on the (101) crystal surface, the orthorhombic space group *Pnma* allows Kerr signals from $\chi^{(3)}_{xxxx}, \chi^{(3)}_{yyyy}, \chi^{(3)}_{xxyy} = \chi^3_{xyyx} = \chi^{(3)}_{xyxy}$ and $\chi^{(3)}_{yyxx} = \chi^{(3)}_{yxxy} = \chi^{(3)}_{yxyx}$ (*40*). While the cubic space group *Pm3m* and allows for $\chi^{(3)}_{xxxx} = \chi^{(3)}_{yyyy}$ and $\chi^{(3)}_{xxyy} = \chi^{(3)}_{xyyx} = \chi^{(3)}_{xyxy} = \chi^{(3)}_{yyxx} = \chi^{(3)}_{yxxy} = \chi^{(3)}_{yxyx}$. The *Pnma* space group applies to CsPbBr$_3$ in its orthorhombic phase, which is the case for all temperatures considered in this work. The *Pm3m* space applies to MAPbBr$_3$ for its room temperature cubic phase. All allowed tensor elements were assumed to have the same magnitude.

Simulations for an electronic response only and without optical anisotropy are shown for various thicknesses in **Fig. S13**. This applies to MAPbBr$_3$ single crystals at room temperature when this material is in the cubic phase. Simulations for an electronic response only and with optical anisotropy are shown for various thicknesses in **Fig. S14**. This applies to MAPbBr$_3$ in its low-temperature orthorhombic phase and CsPbBr$_3$, which is orthorhombic for all temperatures considered in this work. The effect of optical anisotropy on the TKE is very dependent on the azimuthal angle of the crystal. Results are shown for two different azimuthal angles: 0° and 45° angle between crystal axis and probe polarization in **Fig. S14**.

**Fig. S14** shows that the oscillatory features due to propagation effects and static birefringence cannot explain the oscillations observed in low temperature MAPbBr$_3$ single crystals and thin films. To simulate this oscillatory signal, we have to consider both electronic and phonon contributions to $R = R_\text{e} + R_\text{ph}$ alongside static birefringence. The chosen simulation parameters are given in the main text and were chosen in accordance with the Lorentzian fits to our experimental data in **Fig. S8**. The instantaneous contribution to $R$ is $R_{\text{e},0}/R_{\text{ph},0}$ times larger than the phononic contribution, when the respective spectral amplitude of the responses are integrated in the 0 - 10 THz range.

**Figures**

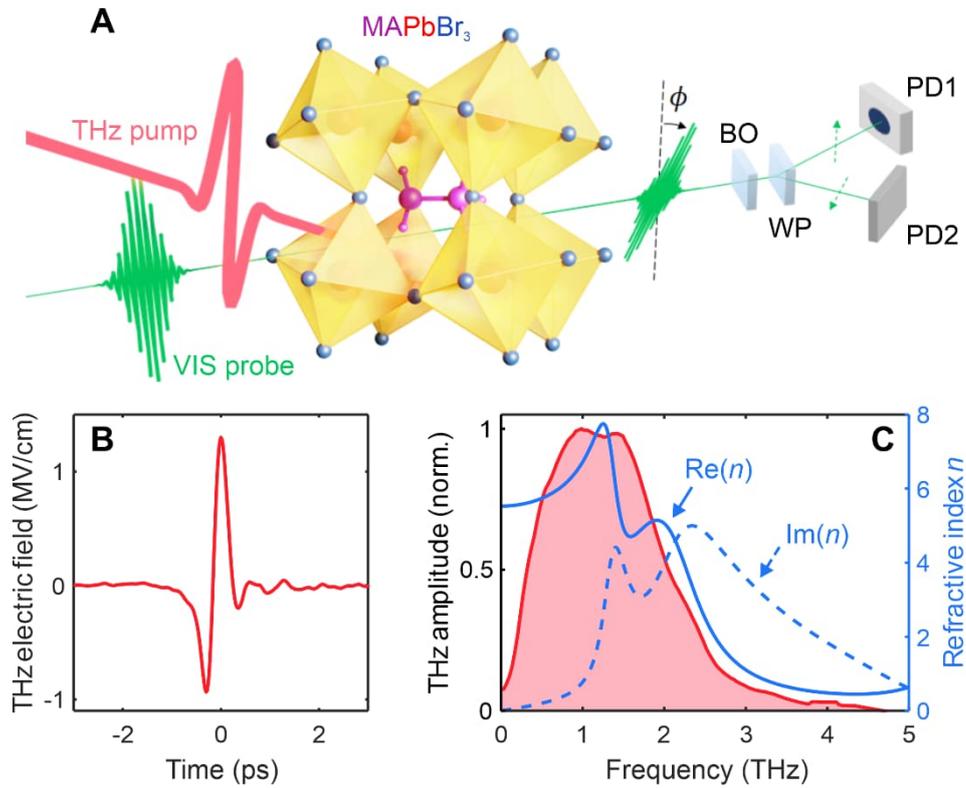

**Fig. 1 | THz fields for nonlinear lattice control in lead halide perovskites. A.** Sketch of the experimental pump-probe configuration. An intense THz electric field causes a transient change of birefringence, leading to an altered probe pulse polarization. This change in polarization is read out using a balanced detection scheme, consisting of balancing optics (BO), Wollaston prism (WP) and two photodiodes (PD1, PD2). **B.** Employed pump THz electric field measured using electro-optic sampling. **C.** Complex refractive index of MAPbBr$_3$ (blue curves) obtained from (*10*) and Fourier transform of THz field (red area) in B.

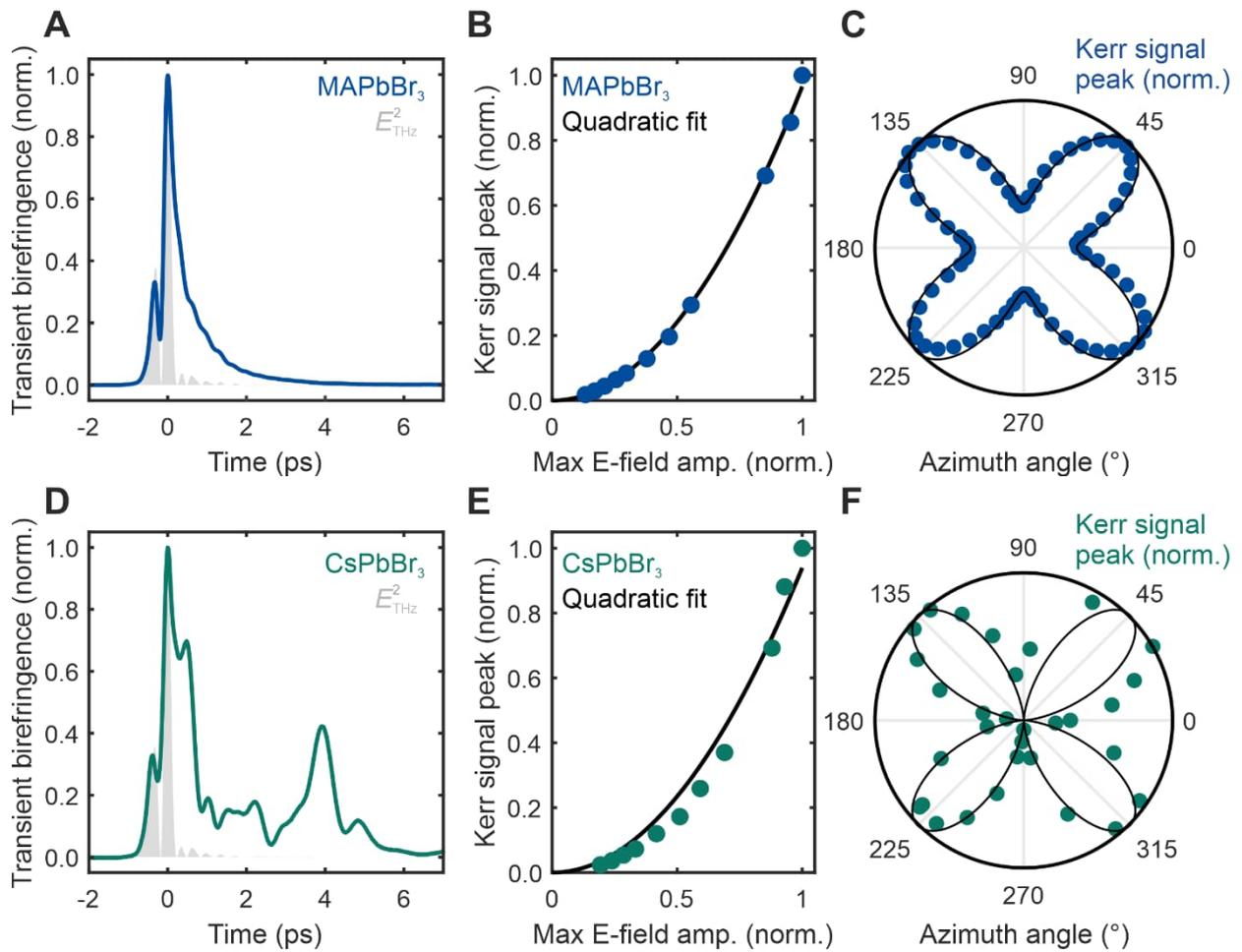

**Fig. 2 | THz-induced birefringence in MAPbBr$_3$ and CsPbBr$_3$ at room temperature. A.** Room temperature TKE in MAPbBr$_3$ and **D.** CsPbBr$_3$ single crystals. **B.** and **E.**, THz fluence dependence of transient birefringence peak amplitude with quadratic fit (black line), demonstrating the Kerr effect nature of the signals. **C.** and **F.**, azimuth angle (between probe beam polarization and crystal facets) dependence of main TKE peak with fit (black line) to expected $\chi^{(3)}$ tensor geometries in cubic and orthorhombic phase, respectively.

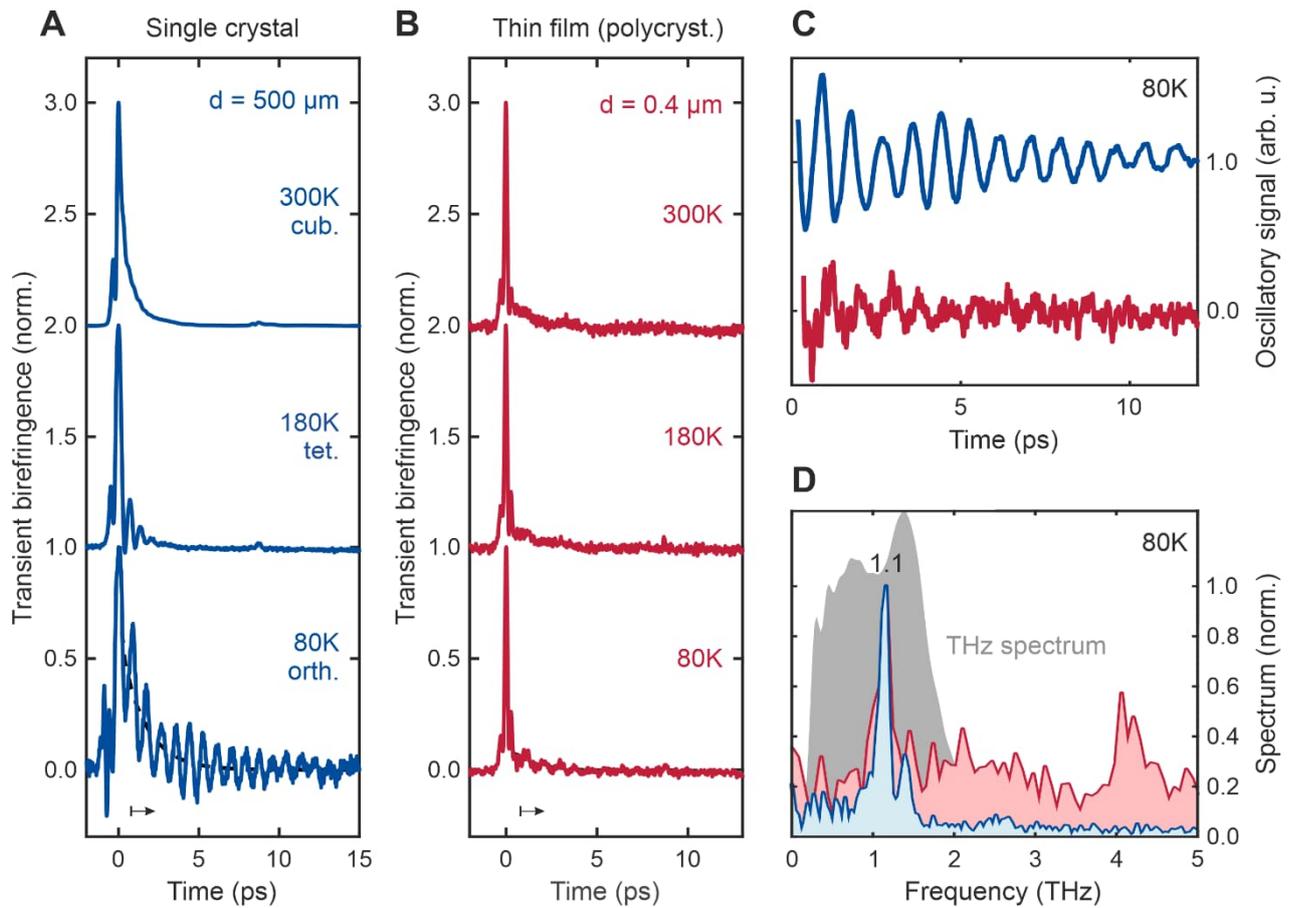

**Fig. 3 | TKE temperature dependence of single crystal vs. thin film MAPbBr$_3$. A.** Temperature-dependent TKE for single crystal and **B.** thin film samples. **C.** Oscillatory signal components at 80K extracted by subtracting an exponential tail (dashed lines) and starting after the main peak (bottom black arrow) in A, B. **D.** Respective Fourier transforms (blue and red) of C and incident THz pump spectrum (gray area).

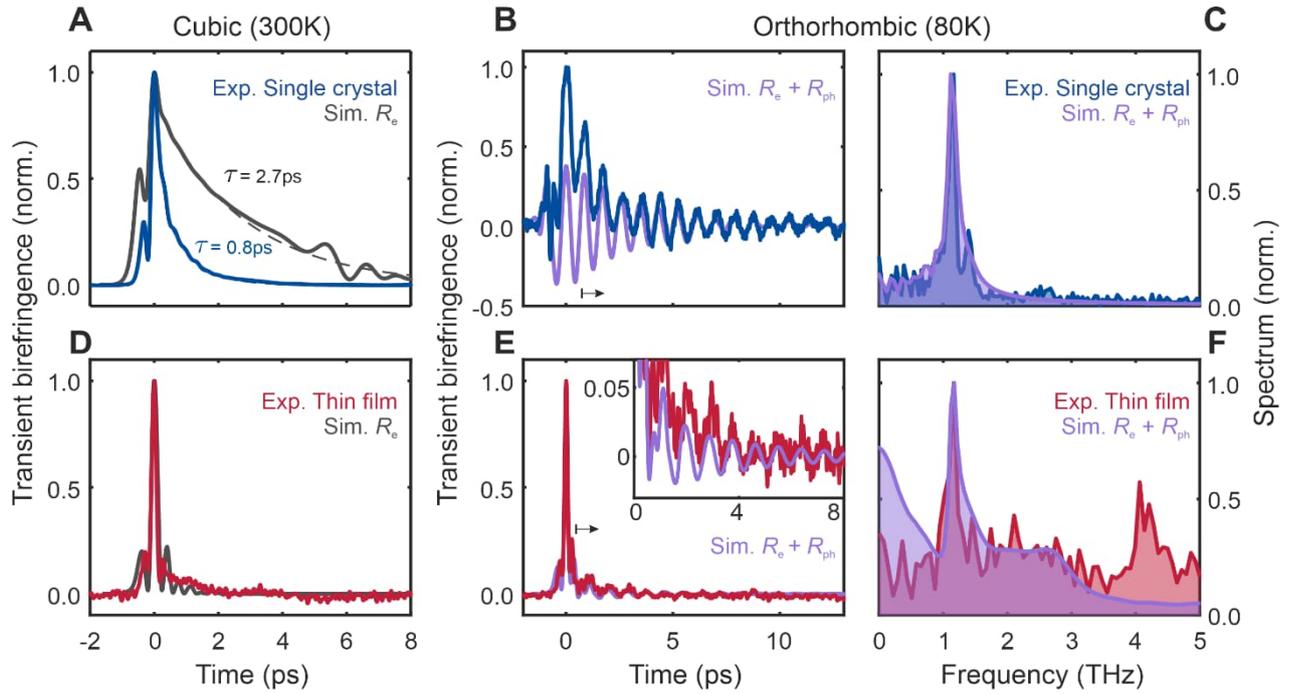

**Fig. 4 | Four-wave mixing simulations vs. experimental results in MAPbBr$_3$.** Isotropic cubic phase (300 K): Simulated TKE signals for **A.** single crystal (500 µm thickness) and **D.** thin film (400 nm thickness) assuming only instantaneous electronic response $R_e(t)$ (gray lines). Anisotropic orthorhombic phase (80 K): **B.** Single crystal and **E.** thin film TKE vs simulation for model system with static birefringence, instantaneous electronic $R_e(t)$ and Lorentz oscillator $R_{ph}(t)$ phonon response (purple lines). **C., F.** Fourier transforms of experimental data (blue and red) and simulation results (purple) from B., E., respectively, normalized to the phonon amplitude at 1.15 THz.

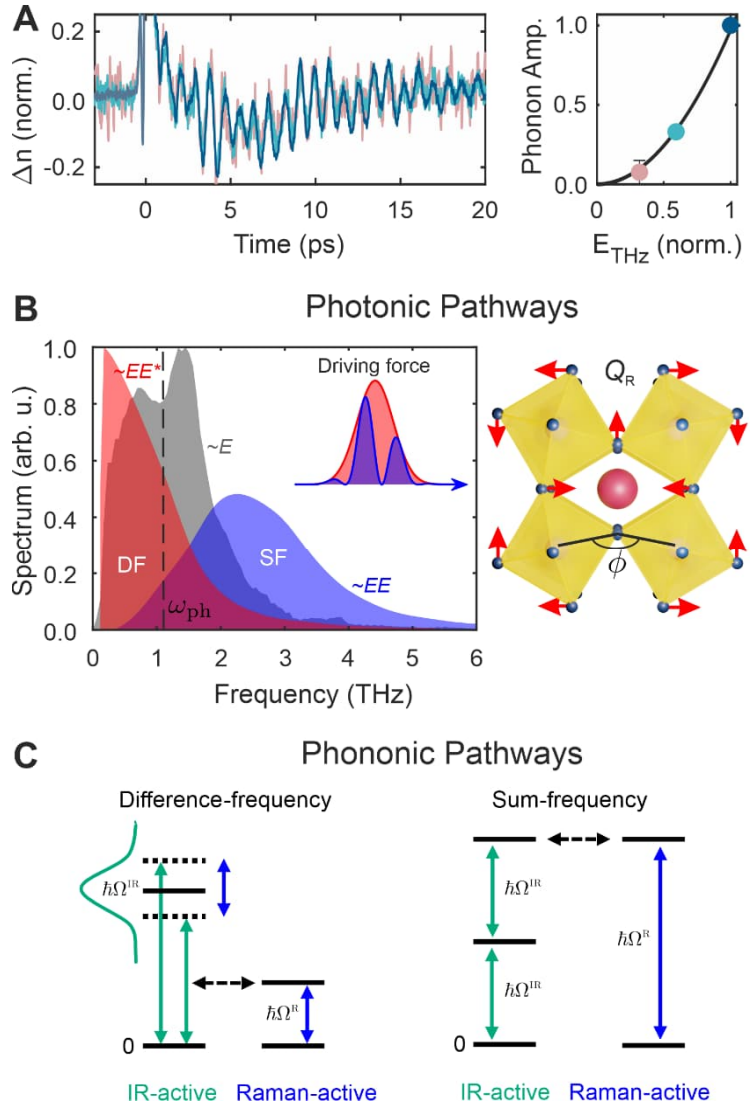

**Fig. 5 | Nonlinear excitation pathways for the 1.15 THz Raman-active twist mode. A.** Time domain coherent phonon oscillations (normalized to $t = 0$ TKE main peak) at 80 K for different THz field strengths (left panel) and respective coherent phonon amplitude (right panel) obtained from Fourier transform; both unveiling a $E_{THz}^2$ scaling law and thus demonstrating a nonlinear excitation. **B.** Possible nonlinear photonic excitation pathways for the $\omega_{ph} = 1.15$ THz mode (dashed line) mediated via a THz electronic polarizability. The nonlinearly coupled $E_{THz}$ spectrum (gray area) leads to difference-frequency $E_{THz}E_{THz}^*$ (DF, red area) and sum-frequency $E_{THz}E_{THz}$ (SF, blue area) driving forces. The octahedral twist mode is schematically sketched on the right hand side. **C.** Possible phononic pathways via a directly driven IR-active phonon $Q_{IR}$, which nonlinearly couples to the Raman-active mode $Q_R$ via anharmonic $Q_R Q_{IR}^2$ coupling.

**Acknowledgments**
We thank A. Paarmann, M. S. Spencer, M. Chergui, A. Mattoni, and H. Seiler for fruitful discussion.

**Funding:** S.F.M. acknowledges funding for his Emmy Noether group from the Deutsche Forschungsgemeinschaft (DFG, German Research Foundation, Nr. 469405347). S.F.M and L.P acknowledge support of the 2D-HYPE project from the Deutsche Forschungsgemeinschaft (DFG, German Research Foundation, Nr. 490867834) and Agence Nationale de la Recherche (ANR, Nr. ANR-21-CE30-0059), respectively. XYZ acknowledges support by the Vannevar Bush Faculty Fellowship through Office of Naval Research Grant # N00014-18-1-2080. M.C. was supported by the DAAD Scholarship 57507869.

**Author contributions:** S.F.M. conceived the experimental idea; M.F., M.C., and S.F.M. designed the research; M.F., M.C., L.N. performed experiments; F.W., B.X. prepared samples; M.F., M.C. analyzed data; J.U., L.H., S.F.M. contributed theory/analytic tools. L.H. developed FWM model and M.F. carried out FWM simulations. M.F., X.-Y.Z., and S.F.M. wrote the manuscript. All authors read, discussed and commented the manuscript. M.F. and M.C. contributed equally to this work.

**Competing interests:** The authors declare that they have no competing interests.

**Data and materials availability:** All data and simulation codes will be uploaded to a public repository after publication of the manuscript.

**Supplementary materials**

**Supplementary information**

1.1 CsPbBr$_3$ TKE temperature dependence

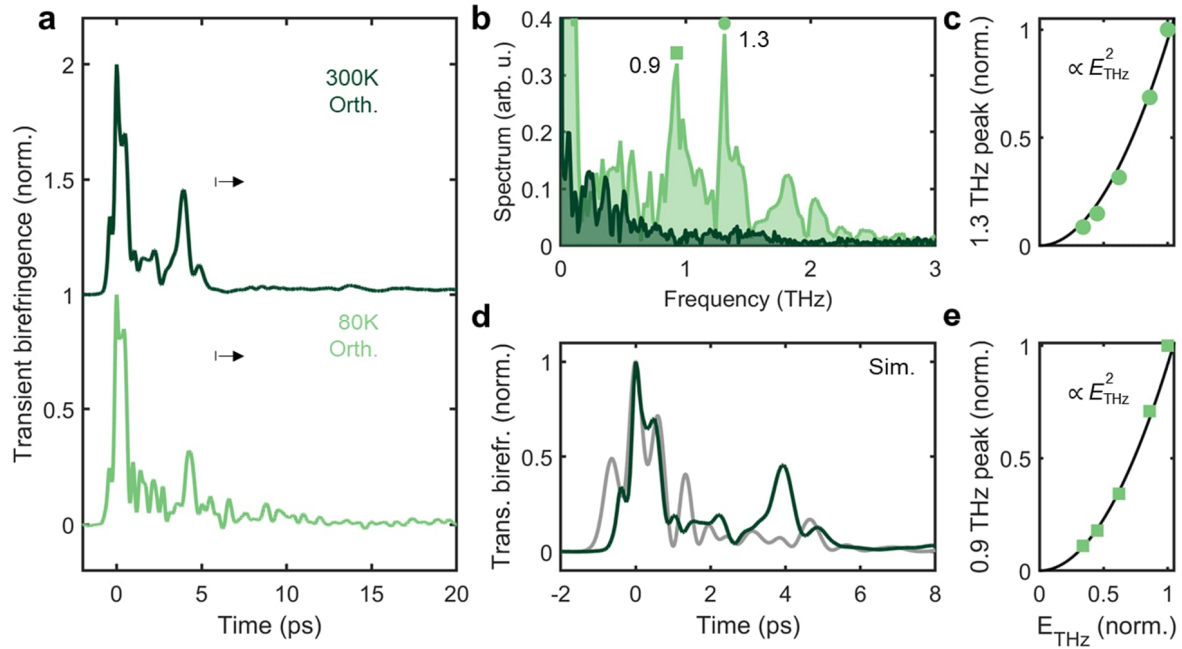

**Fig. S1 | TKE temperature evolution in CsPbBr$_3$. a.** TKE in CsPbBr$_3$ at RT and 80K. In contrast to MAPbBr$_3$, where the structural phase changes for lower temperatures (from cubic to tetragonal to orthorhombic), CsPbBr$_3$ remains in the orthorhombic phase as the temperature is lowered. This is also reflected in the overall TKE shape. However, additional oscillations are visible on the longer timescales at 80K. **b.** Fourier transforming the oscillations after the time indicated by the arrow reveals two main frequency components of 0.9 and 1.3 THz. These frequencies agree well with the two dominating phonon modes in the static Raman spectra of CsPbBr$_3$ (*54*). **c, e.** THz fluence dependence reveals that both oscillation amplitudes (for 0.9 and 1.3 THz) scale quadratically with the THz electric field. **d.** Comparison between simulation for an anisotropic material (100 µm thick and 22.5° azimuthal angle between crystal axis and probe polarization) considering an electronic response only and experimental room temperature CsPbBr$_3$ TKE. This shows that the complex CsPbBr$_3$ TKE signal may be understood in terms of an instantaneous electronic polarization response alongside anisotropic light propagation.

1.2 Estimating the THz nonlinear refractive index of MAPbBr$_3$

Fig. S5 shows a comparison between the TKE in MAPbBr$_3$ and Diamond. The measured TKE signal strength $S(d) = \Delta I/I_0$, where $I_0$ is the total probe intensity measured by the photodiodes and $\Delta I$ is the intensity difference, is proportional to $\Delta n \omega_{\text{pr}} d/c_0$ in Diamond, where $d$ is the sample thickness and $\omega_{\text{pr}}$ is the probing frequency.

This simple relation holds because there is no significant THz dispersion in Diamond. However, due to significant THz absorption and dispersion, this relation does not hold in MAPbBr$_3$ as seen in Fig. S4b. For MAPbBr$_3$, $S(d)$ may rather be approximated by $\frac{\Delta n \omega_{\text{pr}}}{c_0} f(d)$, where

$$f(d) = \int_0^d dz \int_0^\infty d\omega E_{\text{THz}}^2(\omega) \exp(-\alpha(\omega) z) / \int_0^\infty d\omega E_{\text{THz}}^2(\omega).$$

**S1**

Here, $E_{\text{THz}}(\omega)$ is the THz pump spectrum and α is the absorption of MAPbBr$_3$ as extracted from the complex refractive index data in Fig. S11.

Since $\Delta n = n_2 c_0 \epsilon_0 E_{\text{THz}}^2$, we may estimate $n_2$ of MAPbBr$_3$ using:

$$n_2^{\text{MA}} = \frac{S_{\text{MA}}(d_{\text{MA}}) d_{\text{D}}}{S_{\text{D}}(d_{\text{D}}) f(d_{\text{MA}})} n_2^{\text{D}}.$$

**S2**

$n_2^{\text{D}}$ of Diamond has been measured to be $3 \times 10^{-16}$ cm$^2$/W for 1 THz pump and 800 nm optical probing (*52*). Based on $\frac{S_{\text{MA}}}{S_{\text{D}}} = 9.4$, $f(d_{MA} = 500 \, \mu m) = 47 \mu m$, we therefore estimate $n_2^{\text{MA}}$ to be $2 \times 10^{-14}$ cm$^2$/W, roughly 80 times higher than $n_2^{\text{D}}$ for 1 THz pump and 800 nm optical probing.

For comparison, $n_2^{\text{MA}}$ has been previously measured in the near-infrared spectral region using the Z-scan technique (*51*). They found a similar order of magnitude of $n_2^{\text{MA}} = 9.5 \times 10^{-14}$ cm$^2$/W at 1000 nm wavelength.

**Supplementary figures**

Experimental figures

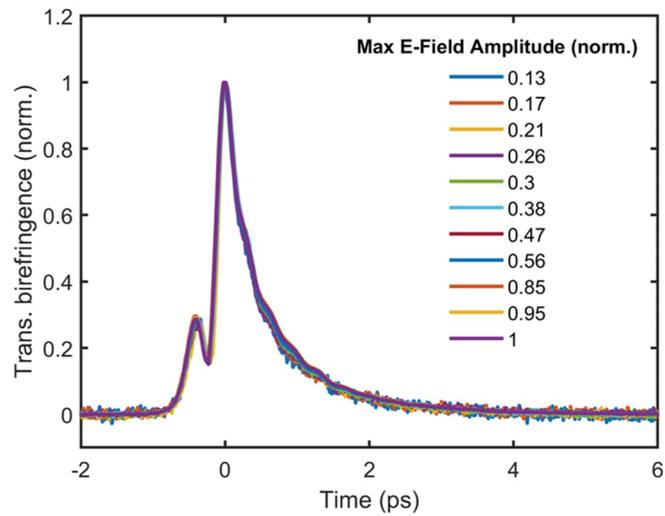

**Fig. S2 | MAPbBr$_3$ TKE temporal dependence on THz fluence.** Normalised experimental TKE of MAPbBr$_3$ at RT for various THz fluences showing that the temporal evolution is not affected by the THz-field strength. Fig. 2 in the main text already showed that the $t = 0$ ps peak scales quadratically with the THz field amplitude.

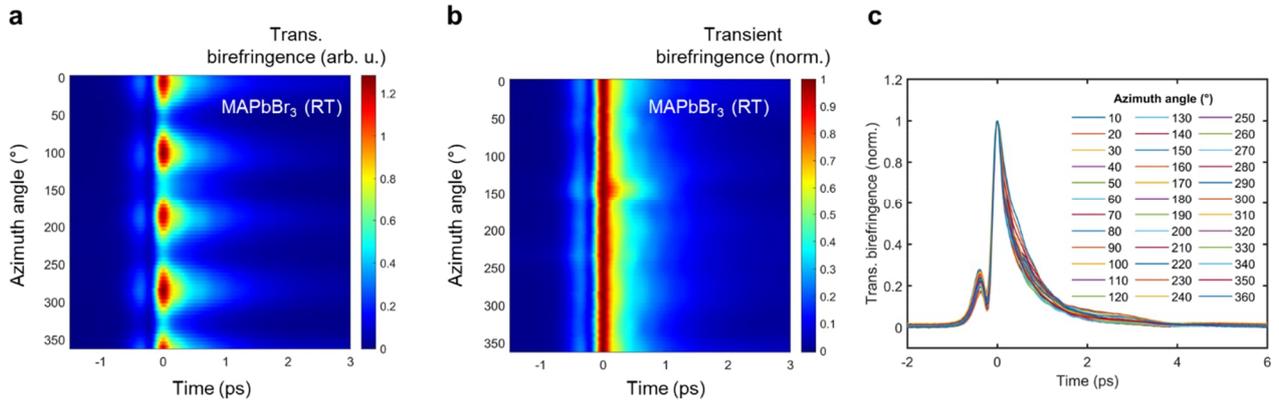

**Fig. S3 | MAPbBr$_3$ TKE azimuthal angle dependence at RT. a.** TKE signal showing the 4-fold rotational symmetry of the measured signal. **b, c.** TKE signal is normalized to show that the time constant of the tail is independent of azimuthal angle. This agrees with the simulations for an isotropic material in Fig. S13, where the origin of the exponential tail is high absorption, dispersion and pump-probe walkoff, which do not depend on the crystal azimuthal angle. Note that the azimuthal angle is not calibrated with respect to the crystal axes in this figure.

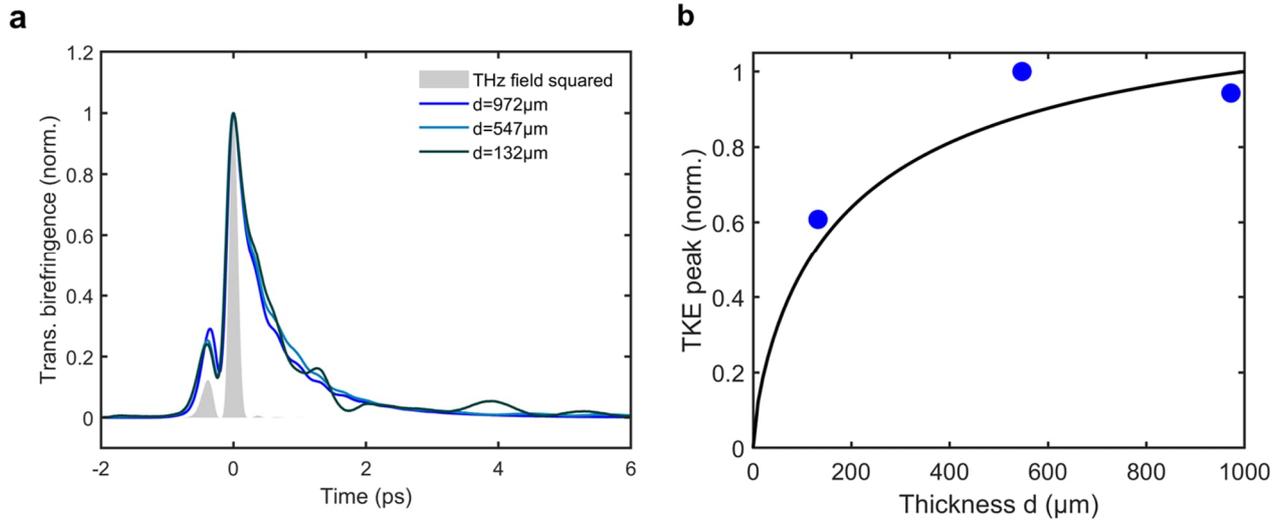

**Fig. S4 | MAPbBr$_3$ TKE dependence on sample thickness. a.** Normalised experimental TKE thickness dependence of MAPbBr$_3$ at RT. The results agree well with the simulations in Fig. S13. **b.** shows the measured TKE peak signal as a function of sample thickness. The black line shows the expected signal dependence when accounting for strong THz absorption and dispersion using the formula $S(d) = \int_0^d dz \int_0^\infty d\omega E_{THz}^2(\omega)\exp(-\alpha(\omega)z) / \int_0^{1000} dz \int_0^\infty d\omega E_{THz}^2(\omega)\exp(-\alpha(\omega)z)$, where $E_{THz}(\omega)$ is the THz pump spectrum and $\alpha$ is the absorption of MAPbBr$_3$ as extracted from the complex refractive index data in Fig. S11.

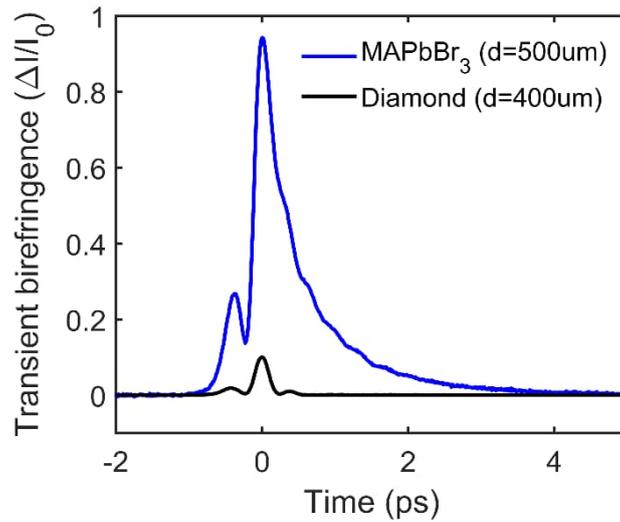

**Fig. S5 | Comparison between TKE in MAPbBr₃ and Diamond for estimating the THz nonlinear refractive index n₂.** Diamond has already been shown to have a strong THz-induced Kerr nonlinearity and be a good nonlinear material in the THz range (*51*). $n_2$ of Diamond has been measured to be $3 \times 10^{-16}$ cm²/W for 1 THz pump and 800 nm optical probing (*52*). For a 500 µm thick MAPbBr₃ single crystal, the TKE peak signal is about 10 times bigger than for a 400 µm thick Diamond.

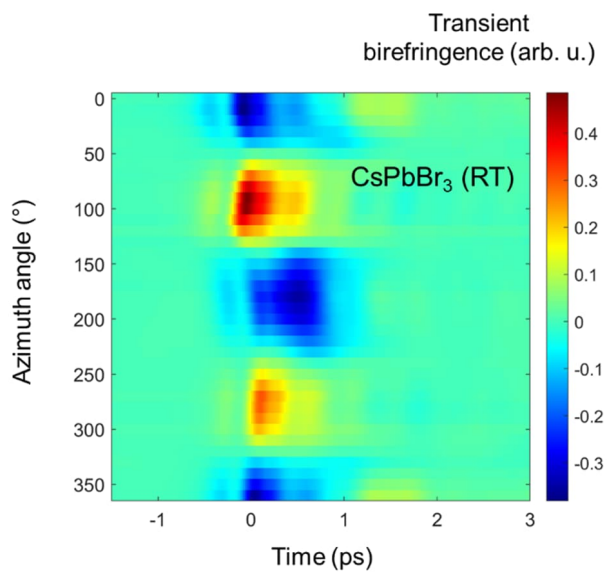

**Fig. S6 | CsPbBr$_3$ TKE azimuthal angle dependence at RT.** Although the main peak exhibits a 4-fold symmetry, the temporal evolution as a function of azimuthal angle is more complex than for MAPbBr$_3$. As CsPbBr$_3$ is in the orthorhombic phase at room temperature, this extra complexity might be explained by additional static birefringence and resulting anistropic light propagation as can be seen in Fig. S14. Note that the azimuthal angle is not calibrated with respect to the crystal axes in this figure.

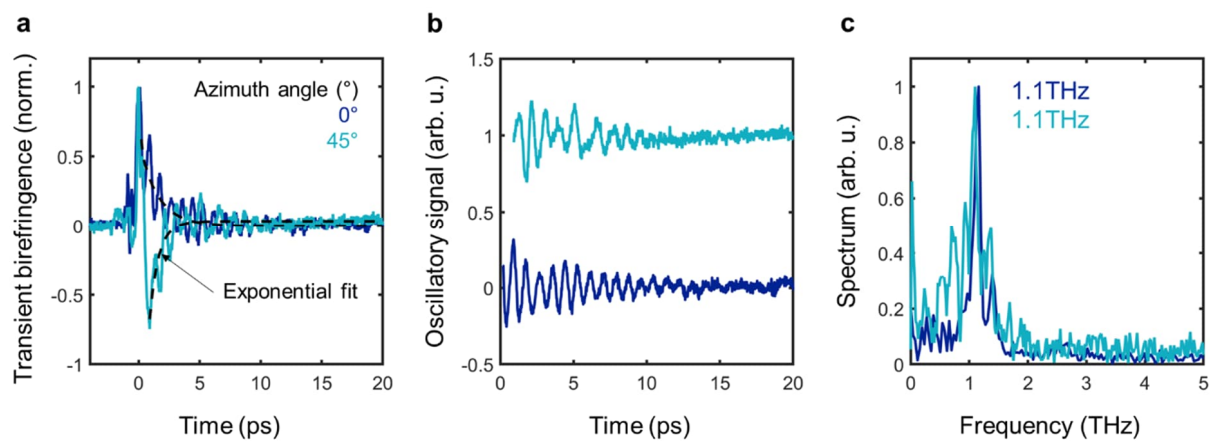

**Fig. S7 | MAPbBr$_3$ TKE at 45° azimuthal angle at 80K. a.** MAPbBr$_3$ TKE at 80K for 0° and about 45° azimuthal angle. MAPbBr$_3$ is orthorhombic at 80K, which might explain the different overall signal shape for both orientations. However, in both TKEs we can see a strong oscillatory signal. **b.** By subtracting off fits to the tails (dotted line in a) for the TKEs at 0° and 45°, we extract the oscillatory signals. **c.** Fourier transforming the oscillatory signals in **b.** reveals that the same 1.1 THz mode dominates the oscillatory response at 0° and 45° azimuthal angle.

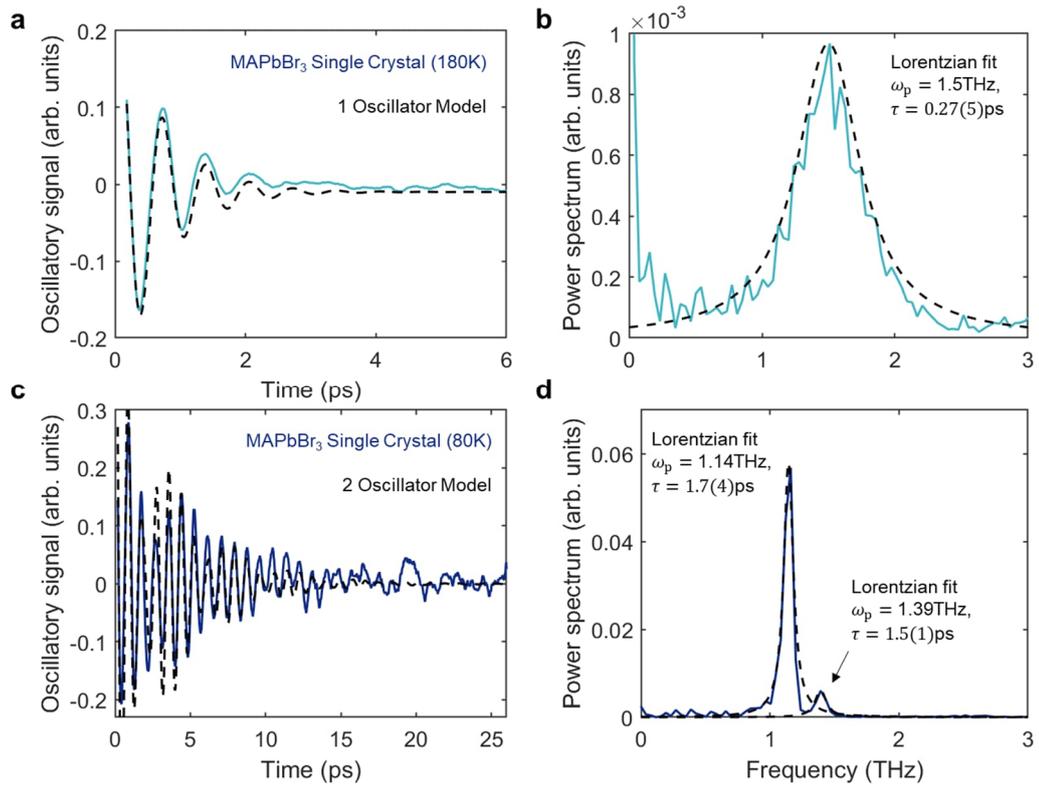

**Fig. S8 | Lorentzian fits to spectral peaks in MAPbBr$_3$ at 180K and 80K. a,c.,** Oscillatory signals extracted from the MAPbBr$_3$ TKE at 180K and 80K in Fig. 3A respectively. The signals are extracted by subtracting off exponential fits to the tails from the main TKE signals. **b.** The modulus squared of the Fourier transform of the oscillatory signal at 180K shows a broad peak at 1.5 THz, which we fit with a Lorentzian. The FWHM of the Lorentzian amplitude is $\Delta\nu_{FWHM}$ =0.58 THz. This corresponds to a phonon lifetime of $\tau = 1/(2\pi\Delta\nu_{FWHM}) = 0.27$ ps. **d.** The modulus squared of the Fourier transform of the oscillatory signal at 80K shows two peaks at 1.14 THz and 1.39 THz. By fitting Lorentzians, we obtain phonon lifetimes of 1.7(4) ps and 1.5(1) ps for the two peaks respectively.

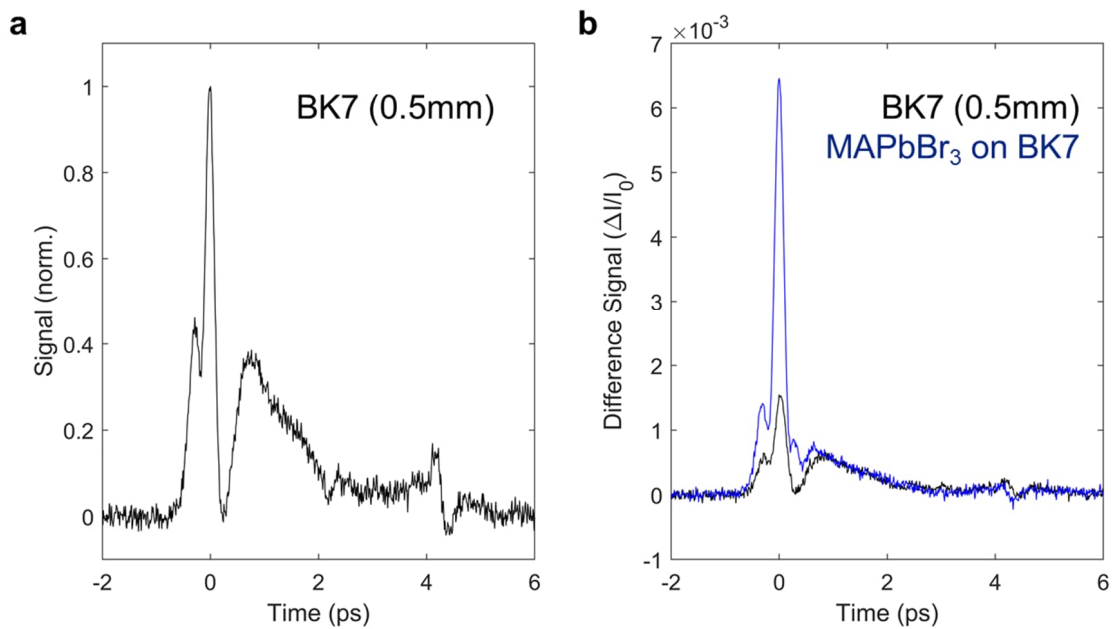

**Fig. S9 | BK7 TKE at room temperature. a.** TKE of a BK7 substrate with 0.5 mm thickness. **b.** shows the BK7 TKE relative to the TKE of a MAPbBr$_3$ thin film on top of a BK7 substrate with 0.5 mm thickness at room temperature.

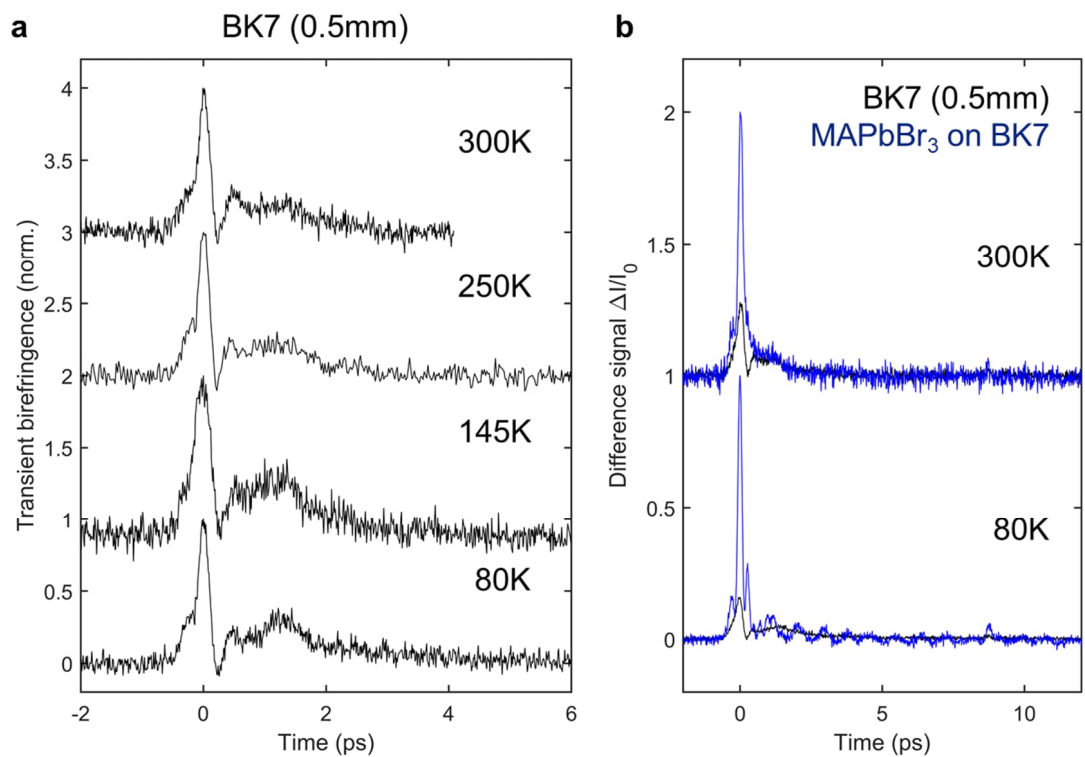

**Fig. S10 | BK7 TKE for various temperatures. a**. TKE temperature dependence of BK7 substrate with 0.5 mm thickness. **b.** shows the relative strength and shape compared to MAPbBr$_3$ thin film on top of a BK7 substrate for room temperature and 80K.

Simulation figures

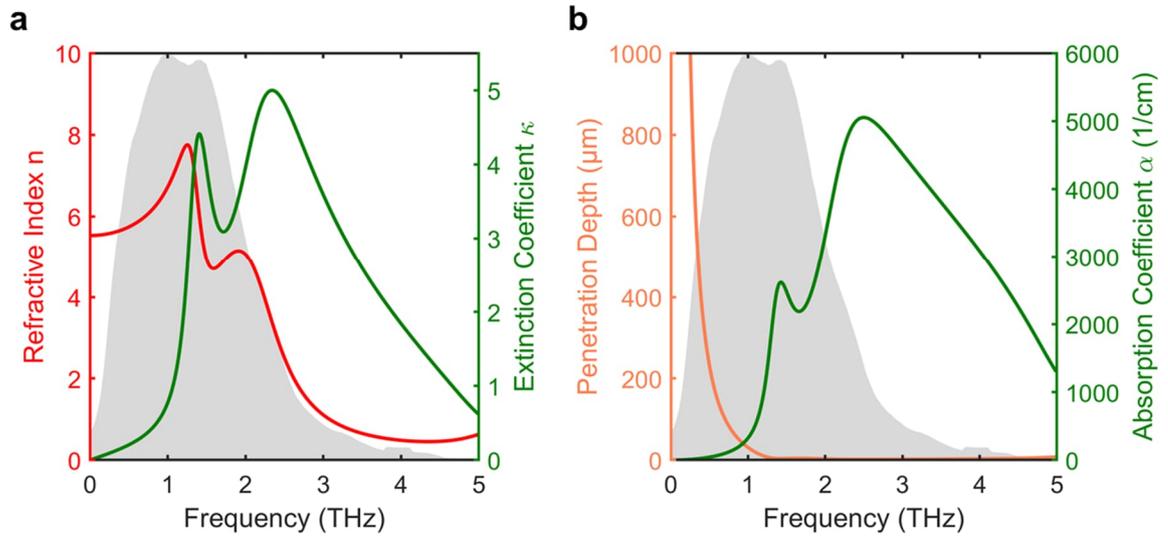

**Fig. S11 | Dispersion of MAPbBr₃ in the THz region. a.** Refractive index $n$ and extinction coefficient $\kappa$ of MAPbBr₃ calculated using the dielectric function from Sendner et al. (*10*). **b.** Absorption coefficient is calculated using relation $\alpha = 4\pi\kappa/\lambda$. The penetration depth is equal to $1/\alpha$.

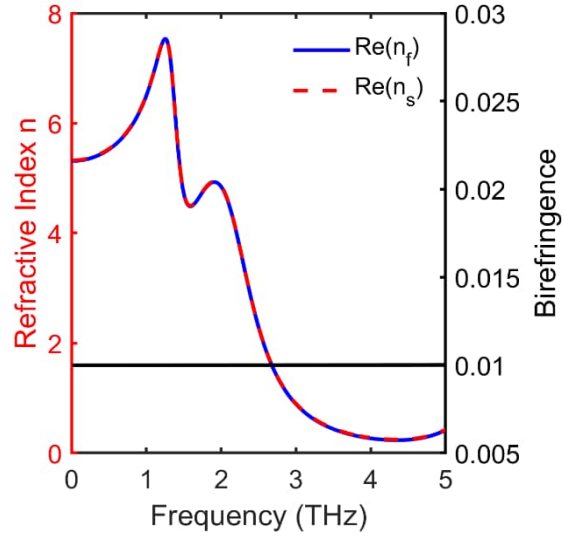

**Fig. S12 | Extrapolated static birefringence of MAPbBr₃ for the simulations of the low temperature orthorhombic phase.** In the optical region, the refractive index of CsPbBr₃ is used as measured using the 2D-OKE (*46*). The static birefringence of CsPbBr₃ is then extrapolated to the THz region. $n_f$ and $n_s$ correspond to the refractive index along the fast and slow crystal axes and static birefringence is defined as the difference between $n_f$ and $n_s$.

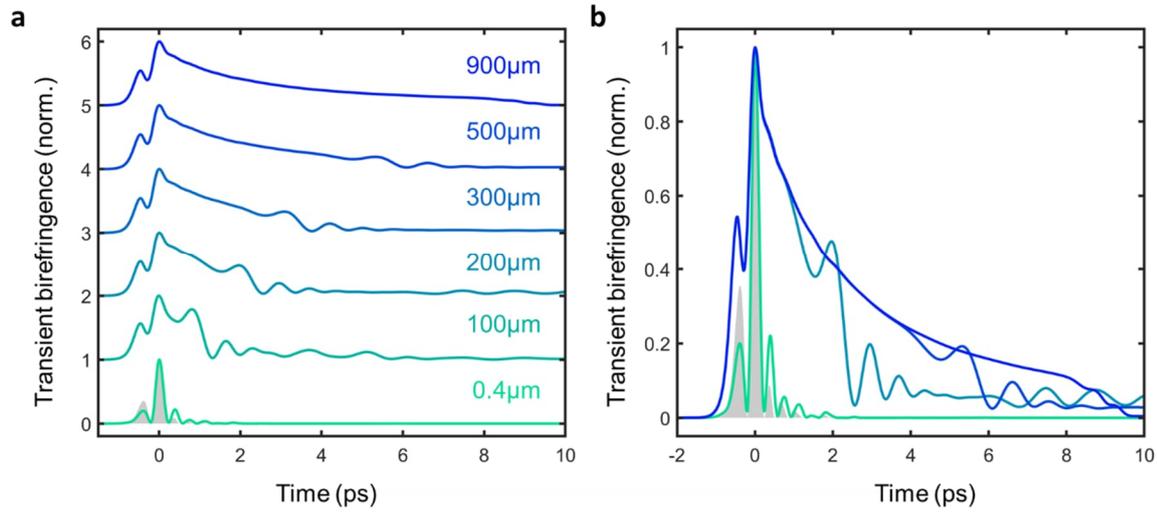

**Fig. S13 | Four-wave-mixing simulation for cubic MAPbBr$_3$ for various thicknesses assuming an instantaneous electronic hyperpolarizability response only. a.** shows the normalised TKE signal for various thickness. For thicknesses larger than 100 µm, we can see an exponential tail with a decay time constant largely independent of thickness. **b**. The normalised TKE signals for various thicknesses are plotted on top of each other. On top of the exponential tail, there are small modulations, whose onset depends on the thickness. The onset time can be roughly estimated by the $t_1$ time ($t_1 = (n_{g,f}(\omega_{THz}) - n_{g,f}(\omega_{pr}))d/c_0$), where $d$ is the sample thickness and $n_g$ is the group velocity refractive index (*40*).

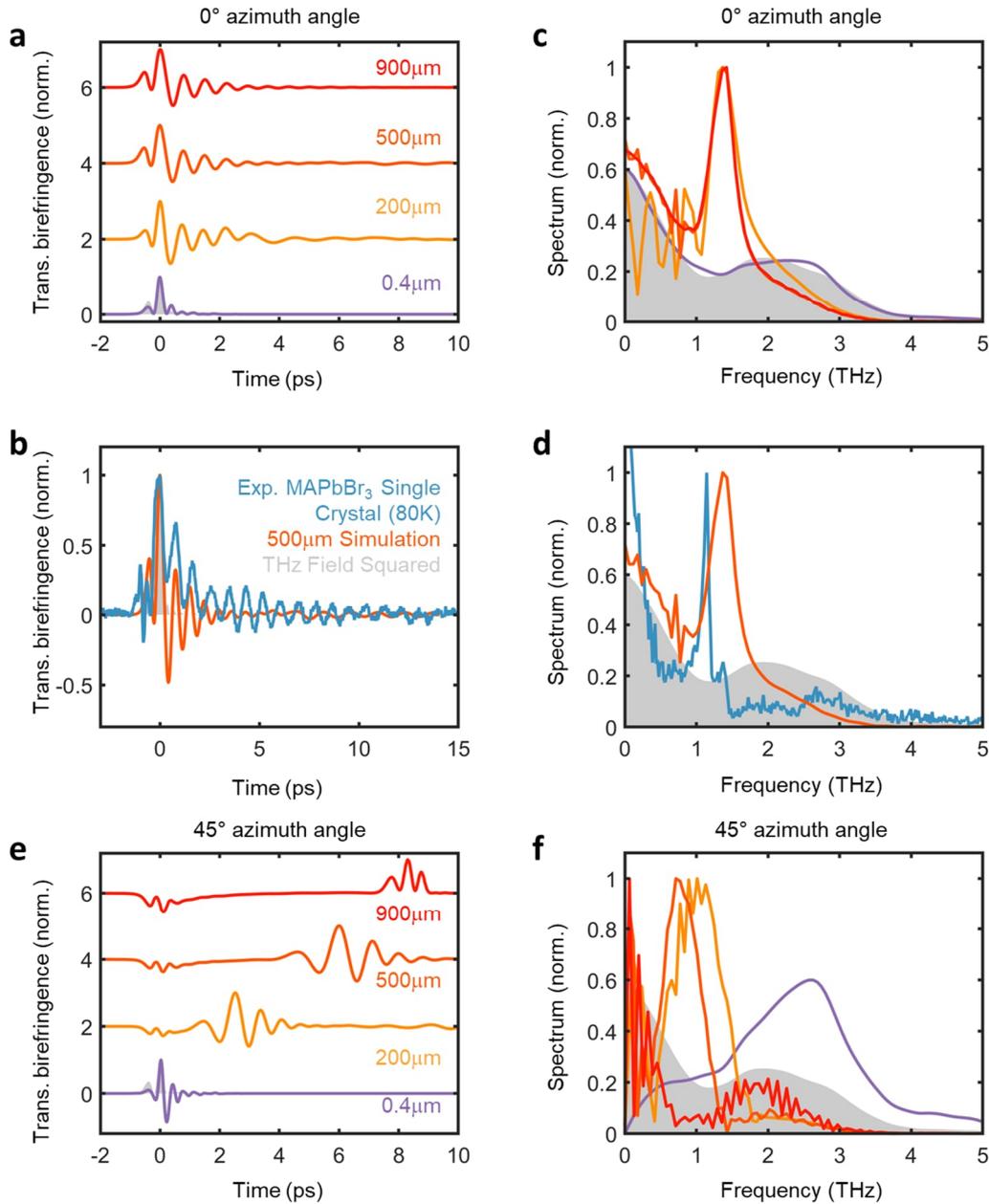

**Fig. S14 | Four-wave-mixing simulation for orthorhombic CsPbBr₃ assuming an instantaneous electronic hyperpolarizability response only.** In contrast to the isotropic simulation in Fig. S13, the azimuthal angle of the model crystal matters for the temporal TKE shape. **a-d**. Results for 0° azimuthal angle of crystal with respect to probe pulse polarization are shown for various thicknesses. For this angle, the birefringence experienced by the probe is maximized. For 0° azimuthal angle and for all thicknesses larger than 200 µm, we can see the appearance of a short-lived oscillatory signal of around 1.4 THz in (**a-d**). These oscillations arise due to static birefringence, but are too short-lived to explain our experimental observation at 80K as shown in (**b, d**). For 0.4 µm thickness, these oscillations due to static birefringence disappear. **e-f**. Results for 45° azimuthal angle for various thicknesses. For this angle, the birefringence experienced by the pump is maximized. The peak t = 0 ps is diminishingly small in comparison to the oscillatory features that happen at later times, which is due to the input tensor symmetry of $R$. The small oscillatory features correspond to internal reflections - similar to the small modulations on top of the tail in Fig. S13. The onset time for these oscillatory features can be roughly estimated by the $t_1$ time.